\documentclass[runningheads]{llncs}
\usepackage{color}
\usepackage{epsfig}
\usepackage{graphicx, subfigure}
\usepackage{amssymb, url}
\usepackage{enumitem, array, tabu}
\usepackage{amsmath}
\usepackage{algorithm}  
\usepackage{algorithmic}  
\usepackage{hyperref}
\usepackage[numbers,sort&compress]{natbib}
\usepackage{listings}
\definecolor{mygreen}{rgb}{0,0.6,0}
\definecolor{mygray}{rgb}{0.5,0.5,0.5}
\definecolor{mymauve}{rgb}{0.58,0,0.82}
\begin{document}
\begin{frontmatter}

\mainmatter

\title{Lost Silence: An emergency response early detection service through continuous processing of telecommunication data streams}

\titlerunning{Lost Silence}

\author{Qianru Zhou\inst{1} \and 
Stephen McLaughlin\inst{1} \and
Alasdair J. G. Gray\inst{2} \and 
Shangbin Wu\inst{3} \and 
Chengxiang Wang\inst{1} 
}
\authorrunning{Zhou et al.} 
%
\tocauthor{Qianru Zhou, Stephen McLaughlin,
Alasdair J. G. Gray, Shangbin Wu, and Chengxiang Wang}
\institute{School of Engineering \& Physical Sciences, Heriot-Watt University, EH14 4AS, UK\\
\and
Department of Computer Science, Heriot-Watt University, EH14 4AS, UK 
\and
Samsung R\&D Institute UK, Staines-upon-Thames, TW18 4QE, UK
}

\maketitle              

\begin{abstract}
Early detection of significant traumatic events, e.g. a terrorist attack or a ship capsizing, is important to ensure that a prompt emergency response can occur. In the modern world telecommunication systems could play a key role in ensuring a successful emergency response by detecting such incidents through significant changes in calls and access to the networks. In this paper a methodology is illustrated to detect such incidents immediately (with the delay in the order of milliseconds), by processing semantically annotated streams of data in cellular telecommunication systems. In our methodology, live information about the position and status of phones are encoded as RDF streams. We propose an algorithm that processes streams of RDF annotated telecommunication data to detect abnormality.  Our approach is exemplified in the context of a passenger cruise ship capsizing. However, the approach is readily translatable to other incidents. Our evaluation results show that with a properly chosen window size, such incidents can be detected efficiently and effectively.

\end{abstract}

\end{frontmatter}

\section{Introduction}
\label{sec_intro}
At 21:30 on 1st June, 2015, the cruise ship ``Eastern Star'' traveling on the Yangtze River began to list as a consequence of stormy weather. One minute later, the ship capsized with 458 passengers and crew on board. No distress signal was sent. It was several hours before the emergency services became aware of the tragedy and 442 lives were lost \cite{peach-time-disaster}.  Although considerable effort has been devoted to the emergency response to natural and man-made disasters, the fatalities and economic cost due to untimely rescue are still significant \cite{missing-migrates, shipwreck1833}. In 2016 alone, 225,665 refugees arrived in Europe by sea, and approximately 2,933 lost their lives due to ship capsizing accidents during illegal migration \cite{missing-migrates}. Most of the capsizes were not noticed until the survivors swam to shore.  
\par
When a ship capsizes and is incapable of communication by radio, is there any way to detect the ship capsizing as soon as it happens?  In a telecommunication system, the location and status information of a phone are stored in the service providers' databases (the specific name of the databases are visiting location register, VLR, and home location register, HLR). When a cell phone is switched off normally (e.g., by pressing power button, or batteries wearing out), it will register its status as ``detached'' to the database. However, if a device is forcefully shut down (e.g., dropped into the water, physically damaged, or the battery removed), or it enters a blind spot with no signal coverage, it does not have time to register its status \cite{forensics2003}. Within a location update period (this varies from 30 minutes to 1 hour, depending on the service provider), the phone is marked as ``unReachable'' in the database. After that period, its status will be changed to ``Detached'', which is the same as in the situation of a normal shutdown. These status signals can be used to infer abnormal events. For example, a large number of phones losing signal abnormally at the same time and location can be used as an indication of a significant event, potentially catastrophic. Taking advantage of this fact, we present \emph{lost silence}, an early detection algorithm that can detect abnormalities. The location update process in the telecommunication system generates large quantities of data in real time. However, in current heterogenous telecommunication system, data from different service providers adopt different local schemas. By annotating the data with RDF/OWL vocabularies, we can present real time phone information as streams of uniformly RDF encoded linked-data. The pattern is processed with continuous SPARQL (C-SPARQL) queries \cite{streamingWorld2009, csparql2009}. This \emph{lost silence} service is a use case for a large project, which aims at building a semantic intelligent knowledge base for the current heterogenous telecommunication system, develop an technology-independent interface, and identify various types of events with semantic web technologies such as SPARQL and rules. 
\par
This paper is structured as follows. Section \ref{sec_scenario} presents our illustrative scenario based on the ``Eastern Star'' shipwreck accident. Section \ref{sec_background} introduces some technical background knowledge of telecommunication networks that are used in our proposal. Section \ref{sec_ontology} presents the ontology we have adopted in our proposed methodology. Section \ref{sec_algorithm} gives details of the algorithms. The evaluation requirements, results and discussions based on our scenario are highlighted in Section \ref{sec_evaluation}. Related work is presented in Section \ref{sec_related} and our conclusions are given in Section \ref{sec_conclusion}.

\section{Motivating Scenario}
\label{sec_scenario}
We outline a scenario based on the cruise ship ``Eastern Star'' capsizing on the Yangtze River, in Jianli, Hubei, China \cite{peach-time-disaster}. In addition to the Yangtze River, the city of Jianli consists of two main parts, a densely populated zone -- Rongcheng town and a surrounding rural area -- Hongcheng village. The detail of population and geographic information is shown in Table. \ref{table_geoData}. 

\begin{table}[ht]
\caption{Geographic data of the city Jianli.}
\label{table_geoData}
\begin{tabular*}{12cm}{@{\extracolsep{\fill}}lll}
\hline
Town & Area ($km^{2}$) & Population\\ \hline
Rongcheng Town & 85 & $ 152, 358 $ \\ 
Hongcheng Village & 254 & 133, 544 \\ 
Water Area & 14.520 & 0 \\ \hline
\end{tabular*}
\end{table}
 In a city with a busy telecommunication datacenter with high volumes of phones (approx. 286,000 phones in the city covering an area of 373.520 $km^{2}$), we have evaluated the algorithm proposed in this paper against the time taken to detect abnormalities, the accuracy of the detection. In particular, we have investigated the tradeoffs between the time taken in generating linked-data streams for the massive amount of data in telecommunication system, and the accuracy of the abnormalities detected. 
\par
It is possible that some situations would reduce the accuracy of the approach described in this paper. For example, although the telecommunication service providers are trying very hard to achieve full signal coverage, there are some regions, especially in rural areas, that have no or limited signal coverage. These areas are known as blind zones. If a large number of phones enter a blind zone, then a large number of phone signals are lost at the same spot. Each service provider keeps a list of its blind zones. Such false alarms could be avoided by a simple comparison to the list before an alert is sent. 

\section{Background Technology}
\label{sec_background}

\subsection{Stream processing}
In the last few years, stream processing has gained a prominent attention in the Semantic Web community \cite{streamingWorld2009, csparql2009,  cqels2014, sparqlStream2013, streamingTheWeb2014}.  The data in real telecommunication system, e.g., system log, SNMP polling, tcpdump, and configuration data, etc., are extremely dynamic. The granularity of SNMP, Syslog and tcpdump update period is in terms of a few seconds, one second and microseconds, respectively \cite{neural2008}. Thus, there will be significant advantages to being able to manage rapidly changing systems at the semantic level. The proposed lost silence application is powered by the Continuous SPARQL (C-SPARQL) engine \cite{csparql2009}, an extension of SPARQL with the ability to query streaming RDF data in real time. 

\subsection{The location register process in mobile telecommunication networks}
\label{subsec_2_1}
To ground our discussion, it is necessary to consider how the telecommunication system maintains records of the location of a phone. In telecommunication networks, phones measure the channel environment and reselect cells every 200ms, and report its location to the cell tower periodically. This period varies from every 30 minutes to every hour, depending on the service providers \cite{lte}. When a phone is turned off normally, it will deregister itself to the network system and its status will be marked as ``detached''. However, if a phone loses its power abnormally, it has no time to deregister. Thus its status in the network system will be ``unReachable''. After the update period, if the network still does not receive update information from this phone, it will register the phone as ``Detached'', in the same manner as if it had been shut down normally \cite{forensics2003, lte}. Thus, when a phone loses signal abnormally, there is a limited time period to identify whether it has been normally shut down or not. Typically this period is of the order of 30 minutes to an hour. The location information of a phone is sent and stored in the telecommunication network servers (namely home location register, HLR or visitor location register, VLR) located in the datacenter of the telecommunication system. Normally, there is one datacenter for one service provider per city, managing the data of all the phones in the city. There are usually more than one service provider in a city, and these service providers usually adopt different schemas to represent the same data. In cases when public security is concerned, some information is shared and uploaded to an authorised third party. For example, in America, as enforced by the government, the location information of every 911 call is automatically obtained and shown to the police \cite{e911}.  

\subsection{Geo-Pixel}
\label{subsec_geoPixel}
Lost silence needs to identify the spatial location where an abnormally high number of phones lose their signals almost simultaneously. Key to the lost silence algorithm is a uniformed spatial segmentation among the data. The current spatial position representation, Universal Transverse Mercator (UTM),  identifies a location by dividing the earth into six zones, and applying a secant transverse Mercator projection in each zone \cite{utm}. However, as the position information in current telecommunication system is represented by traditional longitude and latitude, a considerable amount of computation will be required to convert the traditional position information into UTM. Thus, we propose the concept of \emph{Geo-Pixel} here. It is defined based on the third decimal degree of the latitude and longitude coordination, with the resolution of $0.001^\circ\times0.001^\circ$, or $100m\times100m$. Thus, with \emph{geo-pixel}, phones are aggregated into grid cells based on location. The main task of lost silence algorithm is to detect the \emph{geo-pixels} with abnormally high number of lost signal. 
\par
We made a simplification with regards to \emph{geo-pixel}. At the equator, one third decimal degree of longitude and latitude both cover about 100 meters. However, when moving toward the pole, the distance represented by one degree of longitude degeneralize to zero, while the latitude stays almost the same, for the distance resolution of longitude (east-west distance) depends on the latitude. For example, the distance between one degree of longitude worth up to 111.320 km when latitude is $0^\circ$, and reduces to only 28.920 km when latitude is $75^\circ$. However, most countries with significant mobile phone using populations lie between $75^\circ$ to $-75^\circ$ latitude. Thus, the spatial resolution of \emph{geo-pixel} varies between $100~m\times100~m$ to $100~m\times30~m$, which does not have any impact on the algorithm except that making the result more accurate for some countries at high latitudes. 

\begin{figure*}[!h]
\begin{center}
\includegraphics[width=\textwidth]{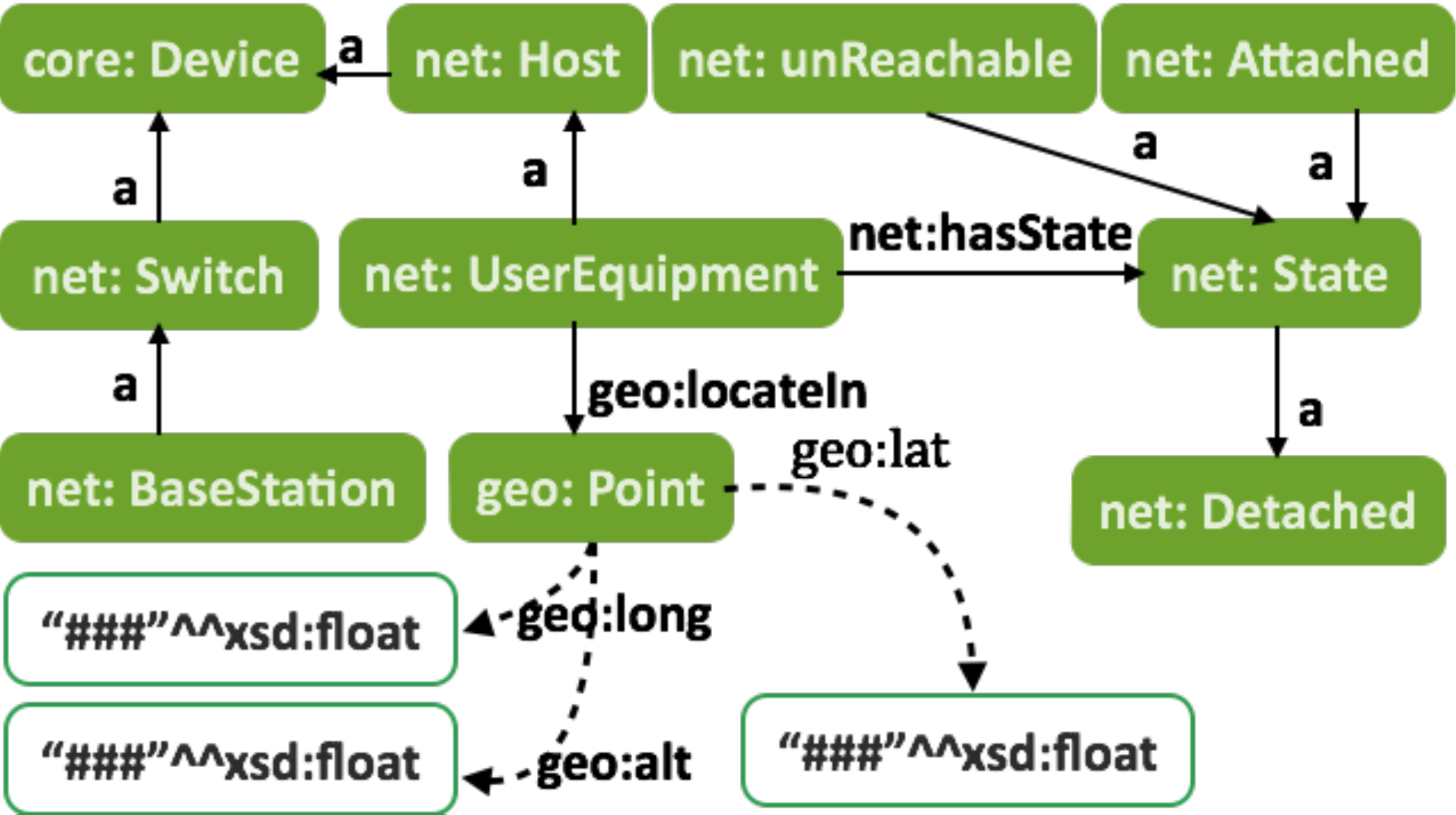} 
\caption{Classes of the TOUCON Ontology used in the Lost Silence scenario. The solid block denotes a class, while the hollow block denotes a data. The solid lines denote object properties, and dash lines denote datatype properties. The prefixes adopted in lost silence are: \emph{net}: $<$\url{http://home.eps.hw.ac.uk/}$\scriptsize{\sim}$\url{qz1/ontologies/wirelessnetwork\_networkResource.owl/}$>$; \emph{core}: $<$\url{http://home.eps.hw.ac.uk/}$\scriptsize{\sim}$\url{qz1/ontologies/wirelessnetwork.owl/}$>$; \emph{geo}: $<$\url{http://www.w3.org/2003/01/geo/wgs84\_pos/}$>$.}
\label{fig_ontology}
\end{center}
\end{figure*}

\section{Ontology adopted in Lost Silence}
\label{sec_ontology}
To be able to transform telecommunication system data into RDF, we adopted an ontology that models the raw data of the system. The ontology which is named TOUCAN Ontology (TOCO) is developed for TOUCAN project\footnote{EPSRC TOUCAN project, No. EP/L020009/1. Website: \url{http://gow.epsrc.ac.uk/NGBOViewGrant.aspx?GrantRef=EP/L020009/1}}. TOCO is proposed to represent the knowledge within a heterogeneous telecommunication system, consisting various technology domains, e.g., mobile network, computer network, optical network, light fidelity (LiFi), and wireless fidelity (WiFi), etc. TOCO is composed of one core ontology and 5 sub-ontologies, namely, \emph{network resource, service, user, time, location}. A diagram of the portion of the TOCO class hierarchy adopted in lost silence is shown in Fig. \ref{fig_ontology}. 
\par

Some important concepts and relations are presented below.
\begin{description}
\item[\emph{``UserEquipment''}] - user equipments in mobile communication system, such as phones, tablets, wearable facilities, etc. It has an object property \texttt{hasStatus} with the object of class \texttt{Status}. It is defined in the network resource sub-ontology. 

\item[\emph{``Point''}] - describes the location of phone, extended from \texttt{SpatialThing} of the W3C WGS-84 vocabulary\footnote{See \url{http://www.w3.org/2003/01/geo/} for more detail.}. It is defined in the location sub-ontology. It has three datatype properties, namely, \texttt{long}, \texttt{lat}, and \texttt{alt}, describing the longitude, latitude, and altitude of the phone.

\item[\emph{``Status''}] - the status of UE connectivity in communication system. There are three instances of this class, namely, \texttt{Attached}, \texttt{Detached}, and \texttt{unReachable}, which denotes the phone is connected, detached after deregister to the system, and unreachable without deregister to the system, respectively. It is defined in the network resource sub-ontology. It is the range of the property \texttt{hasStatus} of \texttt{UserEquipment}.

\end{description}

For example, to describe the fact that \emph{``a phone lost signal at location (329.860, 246.792)''}, the corresponding triples are: \\ 
\emph{net:Phone\_1}\indent\emph{\underline{a}}\indent\emph{net:UserEquipment}.\\ 
\emph{pos:Point\_1}\indent\emph{\underline{a}}\indent\emph{pos:Point}.\\ 
\emph{net:Phone\_1}\indent\emph{\underline{pos:location}}\indent\emph{pos:Point\_1}.\\ 
\emph{pos:Point\_1}\indent\emph{\underline{pos:latitude}}\indent\emph{``329.860''}.\\ 
\emph{pos:Point\_1}\indent\emph{\underline{pos:longitude}}\indent\emph{``246.792''}.\\ 
\emph{net:unReachable}\indent\emph{\underline{a}}\indent\emph{net:Status}.\\
 \emph{net:Phone\_1}\indent\emph{\underline{net:hasStatus}}\indent\emph{net:unReachable}.

\section{Algorithm}
\label{sec_algorithm}
As the data volume from telecommunication system is extremely large, the philosophy of divide-and-conquer is adopted in the algorithm design. In the datacenter of the telecommunication system, for our scenario more than 400 phones lost their signal at the same location within seconds\footnote{By personal correspondence.}. This abnormality could have been detected if continuous queries were being executed on the streams in real-time. An alert could have been raised much earlier and help sent to those on the ship. In the datacenter of the telecommunication system, every time a phone updates its status, a linked-data event is generated and published as a collection of triples. As the phones in the city keep updating their location information and signal status, a linked-data stream is generated. In order to estimate the volume and velocity of data variation from telecommunication system during accidents, we referred to the literatures. The  mobile phones penetration rate is 96.7\% in China, \cite{miit}.  With the \emph{geo-pixel} simplification, the number of geo-pixels and the population density of phones in each town are shown in Table. \ref{table_phonePerGeoPixel}. As a result, in the city of Jianli, there are 29,215 geo-pixels, in which 7,024 of them are in the densely populated zone with on average 21 phones per pixel, 20,991 pixels are in rural areas with on avaerage 6 phones per pixel, and on the river there are 1,200 pixels where the phone density is on average 0 phones per pixel.

\begin{table}[]
\caption{Phone density to geo-pixel in the city Jianli}
\label{table_phonePerGeoPixel}
\begin{tabular*}{12cm}{@{\extracolsep{\fill}}ccc}
\hline
Town & Number of \emph{Geo-pixels} & Phone Density\\ \hline
Rongcheng Town & 7024 &  21  \\ 
Hongcheng Village & 209911 & 6 \\ 
Water Area & 1200 & 0 \\ \hline
\end{tabular*}
\end{table}

\begin{lstlisting}[caption={C-SPARQL query string for detecting the lost phones. Each non-empty result in the returned result sets denotes an alert.}, captionpos=b, label={alg_query}] 
REGISTER QUERY StreamingAndExternalStaticRdfGraph AS
PREFIX xsd: <http://www.w3.org/2001/XMLSchema#>
PREFIX pos: <http://www.w3.org/2003/01/geo/wgs84_pos/>
PREFIX net: <http://home.eps.hw.ac.uk/~qz1/ontologies/ wirelessnetwork\_networkResource.owl/>
PREFIX fn: <http://www.w3.org/2005/xpath-functions#>
SELECT (COUNT(?UE) AS ?counter) ?lat ?long 
FROM STREAM <http://home.eps.hw.ac.uk/~qz1/ontologies/wirelessnetwork_networkResource.owl/stream> [RANGE 30m STEP 5s] 
WHERE { 
 	?UE net:hasStatus net:unReachable. 
	?UE pos:location ?point. 
	?point pos:lat ?lat; 
		pos:long ?long. 
	BIND (fn:round(?lat * 1000) as ?roundLat) 
	BIND (fn:round(?long * 1000) as ?roundLong)
 } 
GROUP BY ?roundLat ?roundLong HAVING (?counter >10) 
\end{lstlisting}

\begin{algorithm*}  
\caption{Stream Generator: Generate RDF streams for each given \emph{geo-pixels}.}   
\label{alg_streamGenerator}  
\begin{algorithmic}
\STATE // Input are the longitude and latitude of the city, ratio of phone lost signal, thread sleep time. Output is the RDF stream created
\REQUIRE $\textbf{\emph{lat, long, lostRatio, sleepTime}}$   
\ENSURE $\textbf{\emph{Stream}}$   
\STATE $keepRunning \Leftarrow 1$   
\WHILE{keepRunning} 
\STATE Insert triple (net:Phone, \underline{geo:locateIn}, geo:Point ) to \textbf{\emph{Stream}}
\STATE Insert triple (geo:Point, \underline{geo:latitude}, ``\textbf{\emph{lat}}''\^{}\^{}xsd:double ) to \textbf{\emph{Stream}}
\STATE Insert triple (geo:Point, \underline{geo:longitude}, ``\textbf{\emph{long}}''\^{}\^{}xsd:double ) to \textbf{\emph{Stream}}
\STATE With a ratio of \textbf{\emph{lostRatio}}: 
\STATE ~~~~~~~~Insert triple (net:Phone, \underline{net:hasStatus}, net:unReunReachableachable) to \textbf{\emph{Stream}}
\STATE With a ratio of$(1 -  \textbf{\emph{lostRatio}})$:
\STATE ~~~~~~~~Insert triple (net:Phone, \underline{net:hasStatus}, net:Attached) to \textbf{\emph{Stream}}
\STATE Thread sleep for \textbf{\emph{sleepTime}} seconds
\ENDWHILE 
\end{algorithmic}  
\end{algorithm*}  

\begin{algorithm*}  
\caption{Thread Pool Scheduler: Schedule the C-SPARQL engine threads.}   
\label{alg_threadPool}  
\begin{algorithmic}
\STATE // Input is the number of \emph{geo-pixels} for city center, rural area, and water zone. Output is the tragedy detection results of the city
\REQUIRE $\textbf{\emph{CityNum, RuralNum, WaterNum}}$   
\ENSURE $\textbf{\emph{Gstream}}$   
\STATE Instantiate a blocking queue: \textbf{\emph{q}}
\STATE Instantiate a thread pool: \textbf{\emph{pool}}   
\FOR{each of the \textbf{\emph{CityNum / RuralNum / WaterNum}} \emph{geo-pixels} in the \textbf{City / Rural area / Water zone}}
\STATE ~~~~~~Create an \textbf{Stream Generator} as a thread into \textbf{\emph{pool}}
\STATE ~~~~~~Apply C-SPARQL query on the stream 
\ENDFOR 
\STATE Shut down \textbf{\emph{pool}}
\end{algorithmic}  
\end{algorithm*}

The C-SPARQL query string to detect lost phones is shown in Listing \ref{alg_query}. One C-SPARQL instance is generated for each \emph{geo-pixel}. Thus, there are 29,215 C-SPARQL instances running in the experiment. With the massive data volume in telecommunication system and the heterogenous population density, it is unrealistic to execute the query based on the arrival of new triples. Thus we choose C-SPARQL, which supports a time step execution model. The query is evaluatgeo-pixeled every 5 seconds calculating for each \emph{geo-pixel} the count of unreachable devices over the last 30 minutes. A result is only returned if the count is above 10. Each stream is denoted as $<$\url{http://home.eps.hw.ac.uk/}$\scriptsize{\sim}$\url{qz1/ontologies/wirelessnetwork/networkResource.owl/stream}$>$ in the query. To avoid the omissions of accidents, we choose the window size the same as the phone's position update period in communication system -- 30 minutes. We adopt \emph{thread-per-geo-pixel} architecture to process telecommunication data and generate a RDF stream in one \emph{geo-pixel} with one thread. The threads are maintained by a thread pool. Each thread in the pool has the same execution priority and receives an equal share of CPU time.  
\par
The details of the generated streams and thread scheduling process are illustrated in an self-explanatory way in Algorithm \ref{alg_streamGenerator} and \ref{alg_threadPool} for brevity. Algorithm \ref{alg_streamGenerator} shows the stream generating process for different areas. The \emph{lostRatio} is adopted to control the ratio of phones that lost signal in the simulated streams. For example, if the $\emph{lostRatio} = 0.1$, it means $10\%$ of the phones in this stream will lose signal, and if the $\emph{lostRatio} = 1$, it denotes all the phones in this stream will lose signal. The density of phones is determined by the \emph{sleepTime}. For example, in the rural area Hongcheng Village where the phone density is 6 phones per pixel, as the location update period is $5~seconds$ ($= 5000~milliseconds$), in order to simulate the scenario that there are 6 phones signal update per $5~seconds$, we have to generate the RDF triples for one phone every $5000 \div 6 \approx 833.333~milliseconds$. Thus, the \emph{sleepTime} for rural area is $833~milliseconds$. Similarly, the \emph{sleepTime} for Rongcheng Town is $5000 \div 21 \approx 238~milliseconds$.

\section{Evaluation}
\label{sec_evaluation}
We run an extensive and exhaustive evaluation based on the tragedy detection algorithm. The evaluation was carried out on a MacBook Air OS X 10.9.5 with 3MB cache running on Intel Core i5 at 1.5 GHz. The system has 128 GB SSD and 4 GB RAM. 

\subsection{Prepare the Data}
A small number of phones will randomly disappear from random pixels along the river. At a randomly chosen time, 424 phones will lose their signal simultaneously in the geo-pixel (329.863, 246.792). With a query window of 30 minutes, we experimented with query steps of 5, 20, and 30 seconds respectively. For the three step sizes, we ran in total 3 iterations of the algorithm. Our evaluation for the shipwreck scenario focus on two criteria: the time taken to detect abnormalities and the accuracy of the detection, in terms of the cases of fail-to-report.
\subsection{Experiment Framework}
\begin{figure}[h]
\begin{center}
\includegraphics[width=3in, height=2.3in]{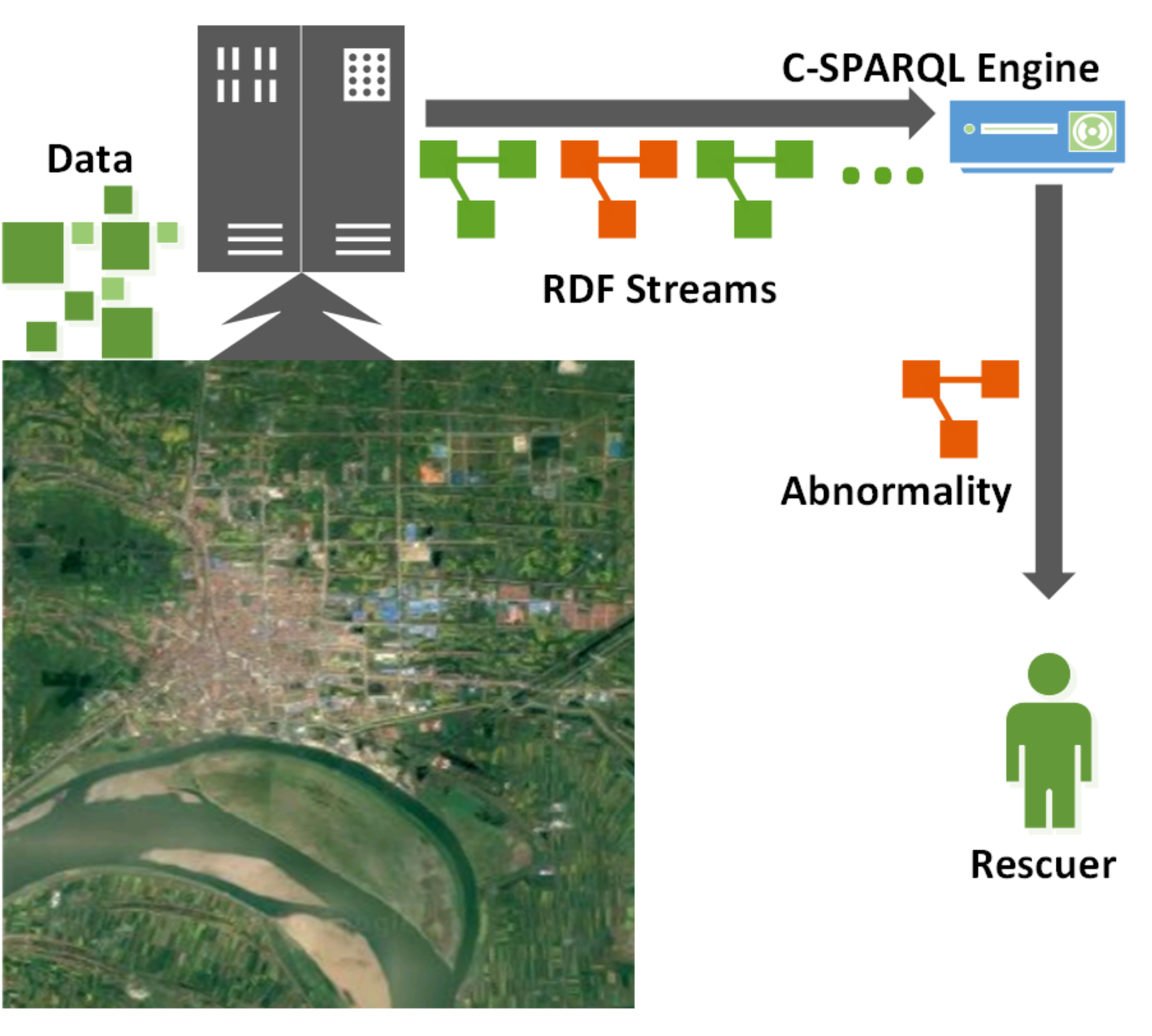} 
\caption{The Process of disaster early detection in lost silence.}
\label{fig_map}
\end{center}
\end{figure}

Fig. \ref{fig_map} shows our workflow and execution environment of the lost silence experiment. Data from the telecommunication system in each \emph{geo-pixel} is converted into RDF streams by the Stream generator.  RDF streams are accessed and queries by the C-SPARQL engine thread. All the 29,215 query threads processing 285,902 phone data are scheduled and executed by thread pool. 

\begin{figure} 
\centering 
\subfigure[]{ \label{fig:subfig:a} 
\includegraphics[width=2.2in]{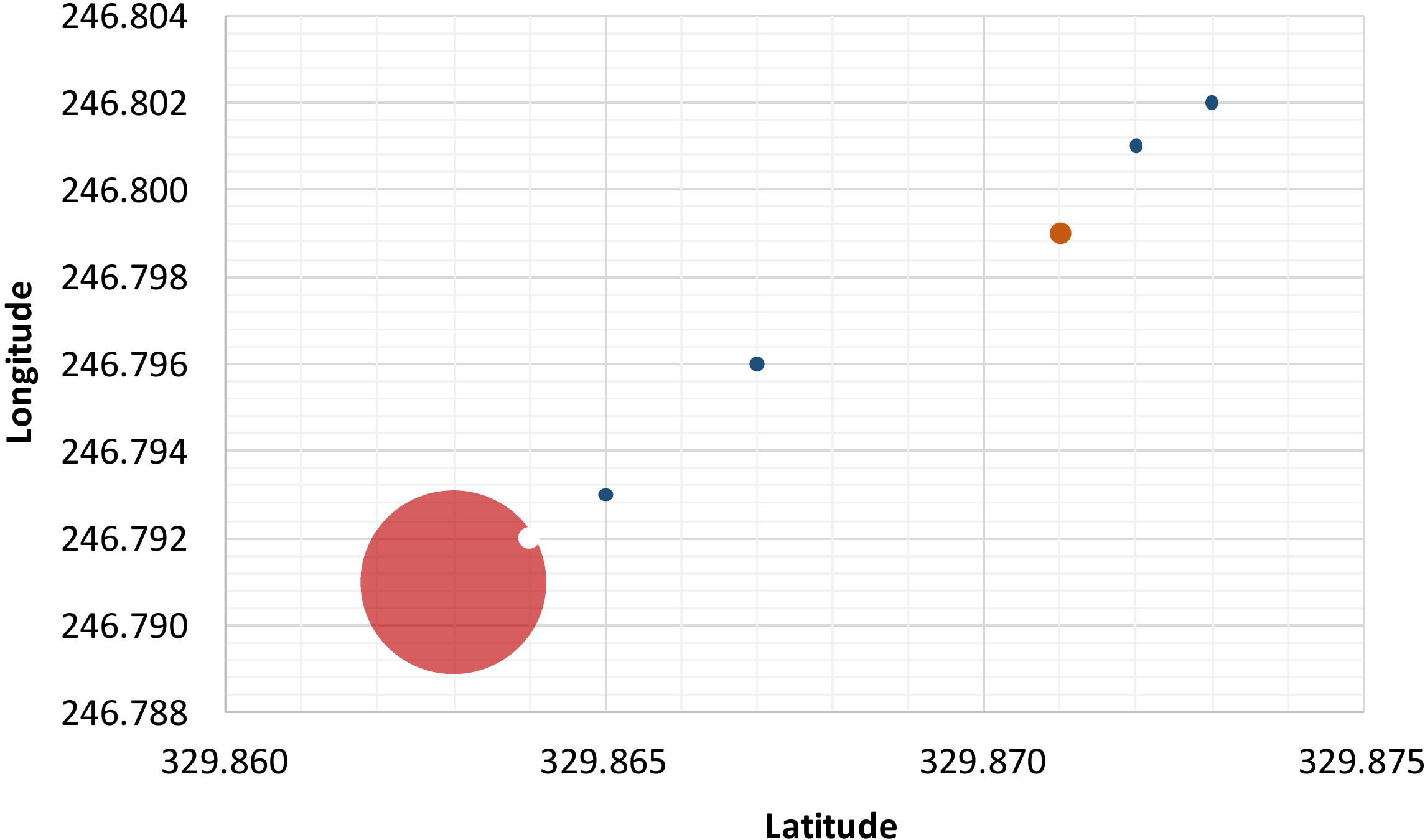}
} 
\subfigure[]{ 
\label{fig:subfig:b} 
\includegraphics[width=2.2in]{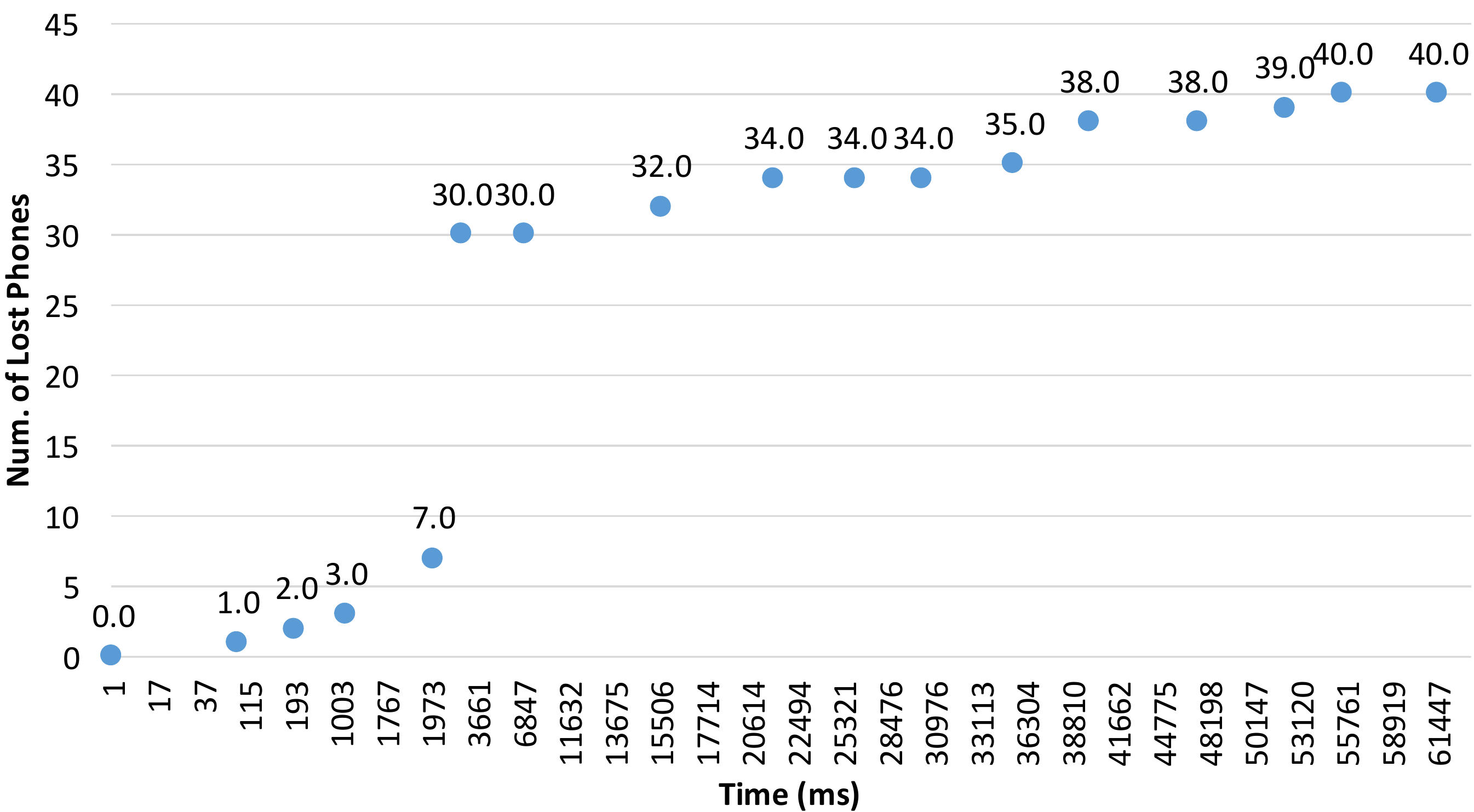}
} 
\subfigure[]{ 
\label{fig:subfig:c} 
\includegraphics[width=2.2in]{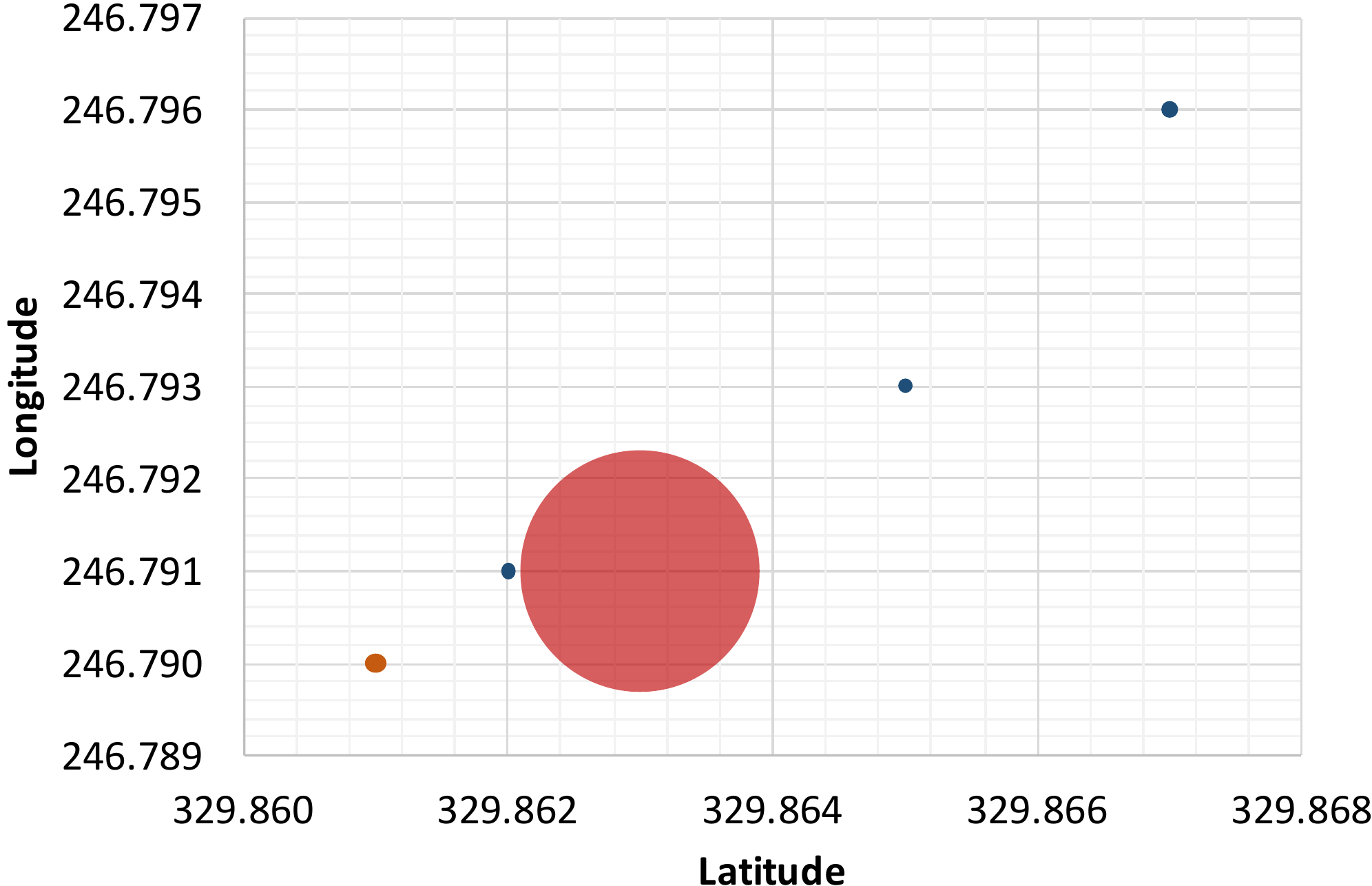}
} 
\subfigure[]{ \label{fig:subfig:d} 
\includegraphics[width=2.2in]{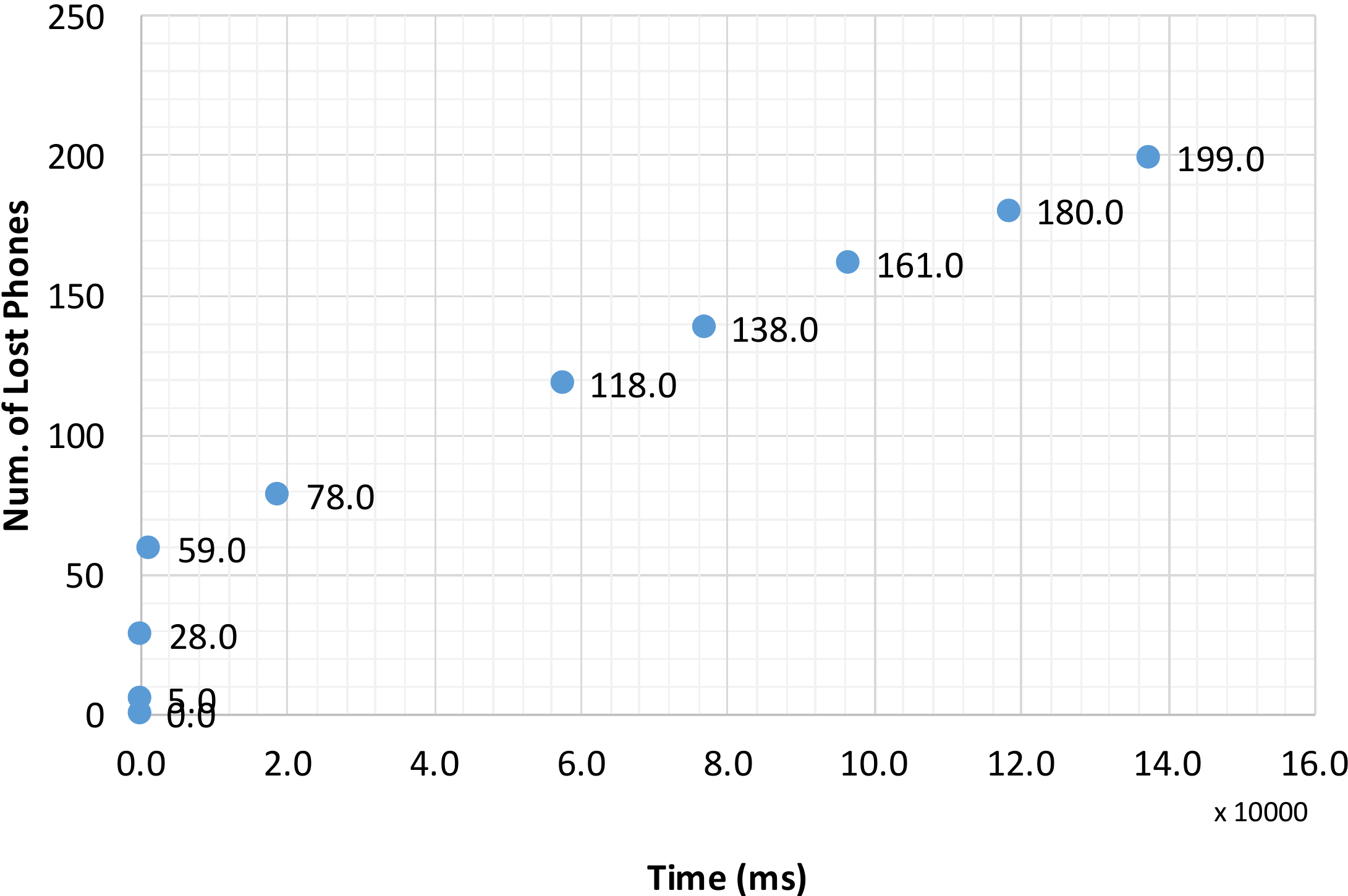}
} 
\subfigure[]{ \label{fig:subfig:e} 
\includegraphics[width=2.2in]{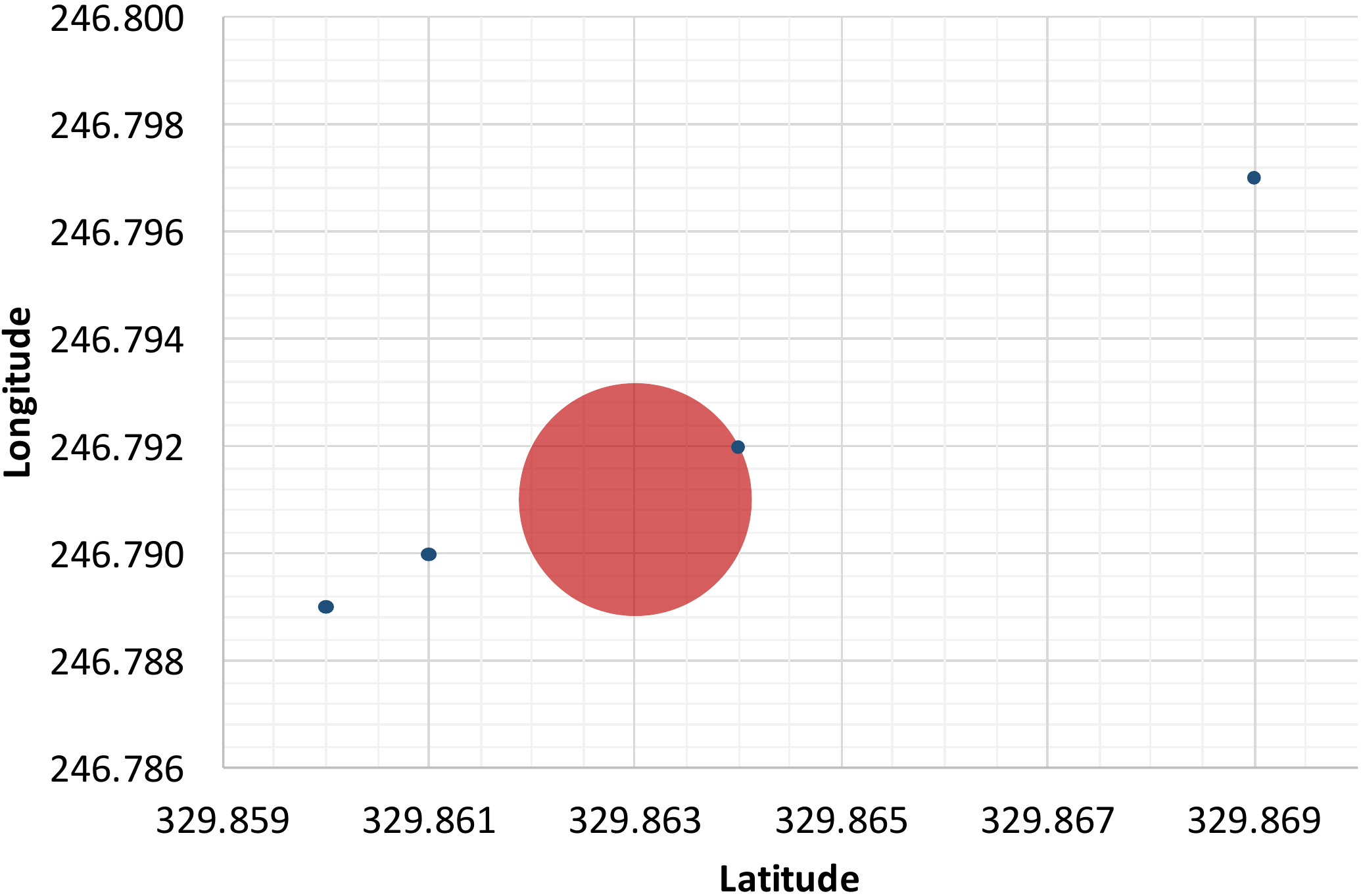}
} 
\subfigure[]{ \label{fig:subfig:f} 
\includegraphics[width=2.2in]{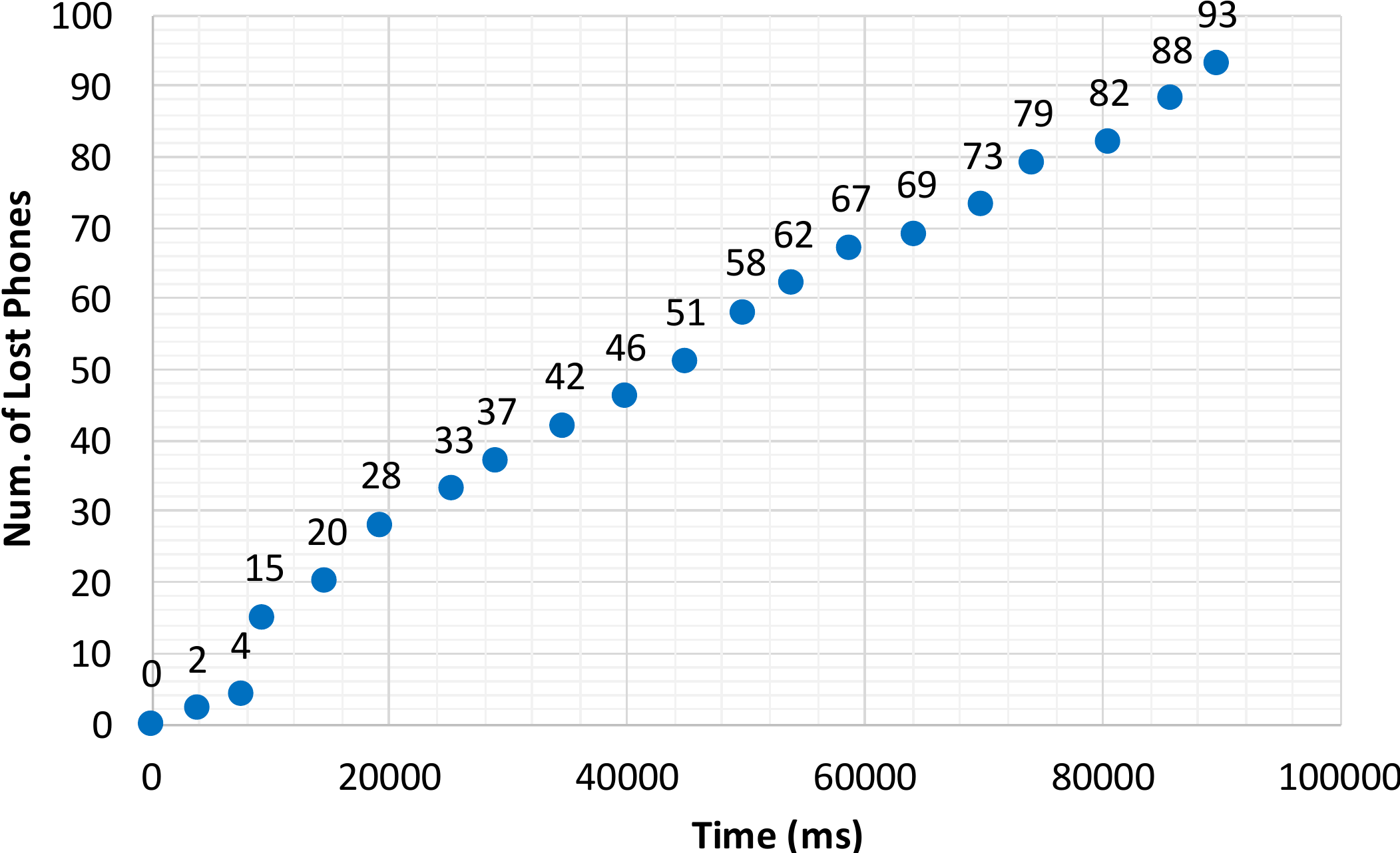}
} 
\caption{The total number of lost phones in the shipwreck area and the query results in the \emph{geo-pixel} (329.863, 246.792), where the shipwreck took place, at each query step for the three experiments with the query steps of 5 seconds, 20 seconds, and 30 seconds, respectively. Fig. a) and b) correspond to the experiment with the query step of 5 seconds, Fig. c) and d) correspond to 20 seconds, and Fig. e) and f) 30 seconds. Fig. a), c), and e) illustrate the total number of phones lost signal at the water area of city Jianli at each experiment. The number of phones lost signal scales with the colour and radius of the circle, e.g., the brighter colour and larger radius of a circle denotes a larger number of lost phones. Fig. b), d), and f) show the number of detected lost phones in the \emph{geo-pixel} (329.863, 246.792), at each query step from the beginning of the query.} 
\label{fig_evaluation} 
\end{figure}

\begin{figure*}[htbp]
\centering
\subfigure[]{ \label{fig_test:subfig:a} 
\includegraphics[width=2.2in]{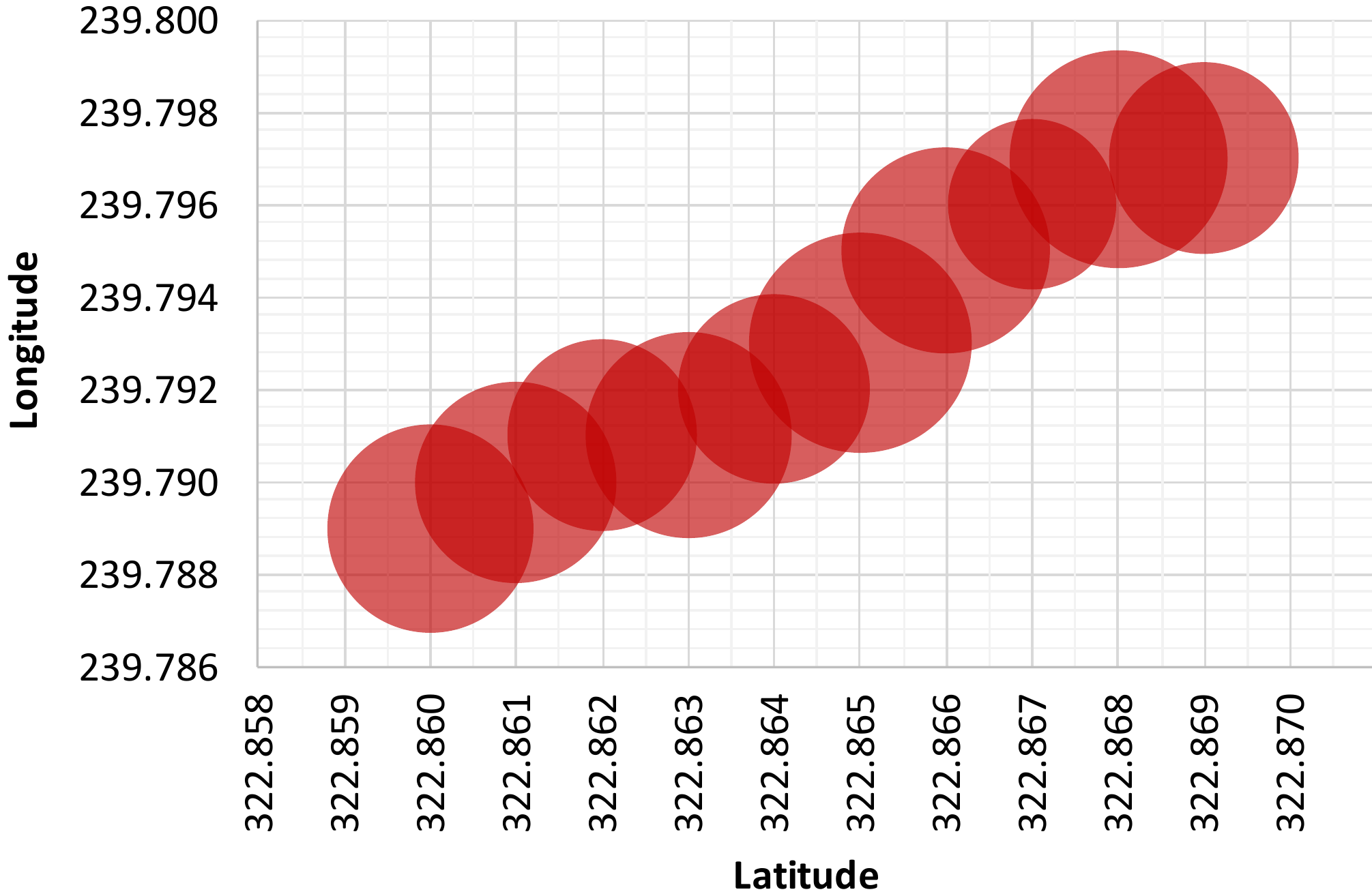}
} 
\subfigure[]{ \label{fig_test:subfig:b} 
\includegraphics[width=2.2in]{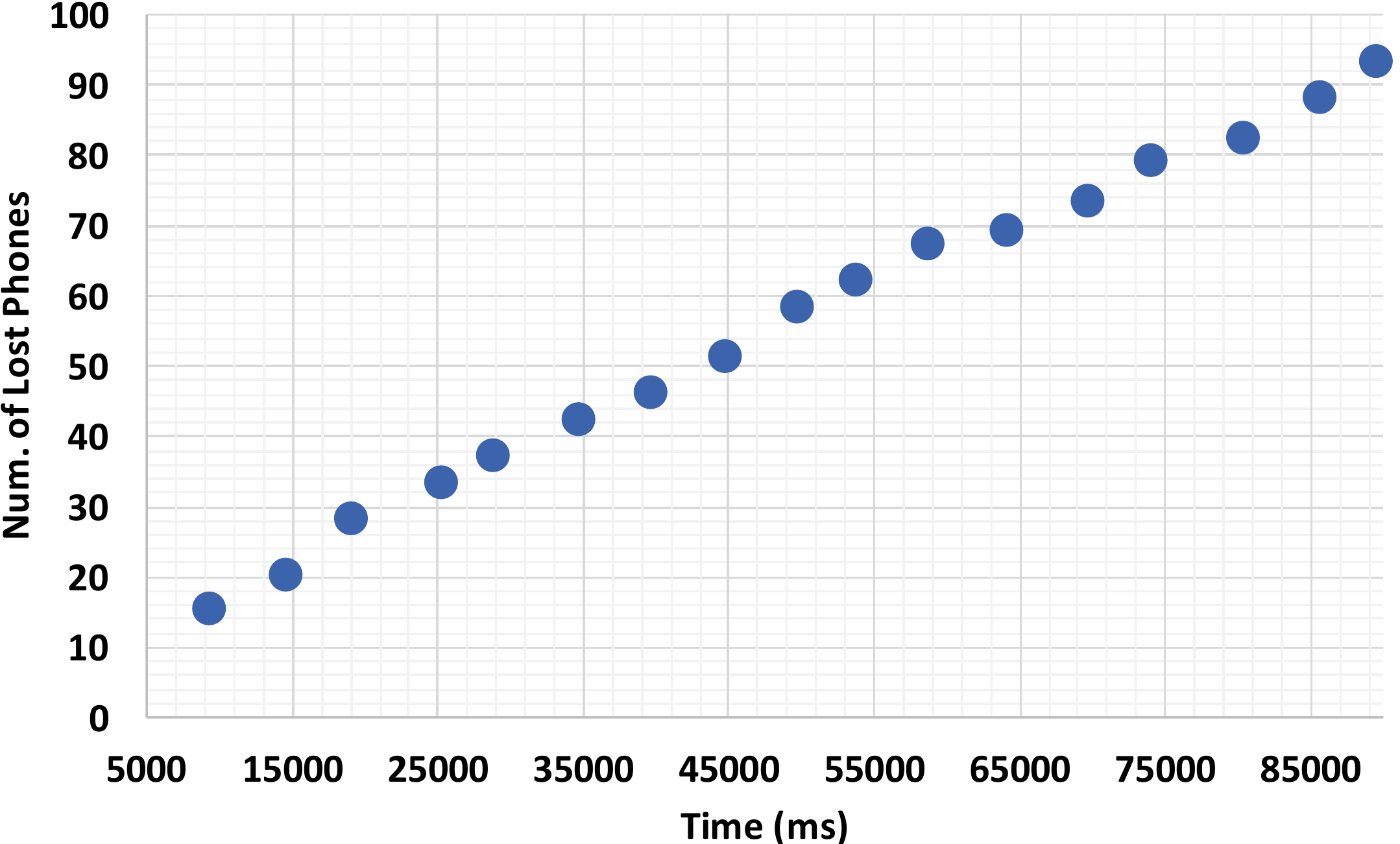}
} 
\caption{Query result in the pressure test scenario in which massive phones lost signal in multiple \emph{geo-pixels} along the river.}
\label{fig_test}
\end{figure*}

\subsection{Evaluation Results}
\label{subsec_evaluationResult}
We run the experiment for three times, with the with query steps of 5 seconds, 20 seconds, and 30 seconds, respectively. The simulation results are shown in Fig. \ref{fig_evaluation}. The Fig. \ref{fig:subfig:a} and \ref{fig:subfig:b} are the simulation result with the query step of 5 seconds, the Fig. \ref{fig:subfig:c} and \ref{fig:subfig:d} are with the query step of 20 seconds, and the Fig. \ref{fig:subfig:e} and \ref{fig:subfig:f} 30 seconds. In the Fig. \ref{fig:subfig:a} \ref{fig:subfig:c}, and \ref{fig:subfig:e} on the left, the horizontal and vertical coordinates indicate the latitude and longitude of the geographical area respectively. The total number of lost phones detected at each \emph{geo-pixel} are shown as circles on the subfigures. The number of lost phones are denoted as the radius and colour of the circles, e.g., the smallest circle in black denotes only one lost phone at that location, the second smallest circle in orange denotes two lost phones, and the largest circle in red is the total 442 lost phones detected where the shipwreck took place. The Fig. \ref{fig:subfig:b}, \ref{fig:subfig:d}, and \ref{fig:subfig:f} on the right show the number of lost phones detected in the geo-pixel (329.863, 246.792), where the shipwreck took place, at each query step, from the beginning of the experiment. To shade light on the detailed detection process, we show the number of lost phones below 10 as well, although no alert will be send for that. As shown in the Fig. \ref{fig:subfig:b}, \ref{fig:subfig:d}, and \ref{fig:subfig:f}, there are sudden jumps of the number of lost phones in each figures, at the time when the shipwreck happens. As the number of lost phones raised to above 10, the lost silence will send alert of that abnormally rise together with the information of the geo-pixel. As the query step grow from 5 seconds (as shown in Fig. \ref{fig:subfig:b}) to 30 seconds (as shown in Fig. \ref{fig:subfig:f}), the number of lost phones increase faster, and the dots intervals are more sparse.
\par
A pressure test is also carried out to simulate an extreme scenario in which 10 ships capsize in different locations on the river simultaneously, as shown in Fig. \ref{fig_test}. Each ship carries more than 300 passengers. The duration of the capsizing of one ship is about 3 minutes. Thus, in the 10 \emph{geo-pixels} along the river, there is a massive phone loss incident with the rate of about $300 \div 3 = 100$ phones per minute. As shown in the  Fig. \ref{fig_test:subfig:a}, the abnormality of a large number of lost phones in all of the 10 \emph{geo-pixels} are detected, and alerts are sent successfully. The results at each query step for an random \emph{geo-pixel} is shown in the Fig. \ref{fig_test:subfig:b}. A sharp increase of the lost phones can be spotted, with the rate about $100$ phones per minute. The result demonstrates the efficiency and practicality of the lost silence. 

\section{Related Work}
\label{sec_related}
The utility of data from telecommunication system in geography and social science to improve urban planning has been increasingly investigated \cite{forensics2003, smartCity2015, citySensing2014, mobileAgnets2006, mobilityPredictability2010}. In \cite{forensics2003}, Willessan presented how evidence obtained from mobile system plays a part in forensic investigation. \cite{smartCity2015, citySensing2014, mobileAgnets2006, mobilityPredictability2010} provided a extensive coverage of the smart city applications adopting data from telecommunication system. However, these application mainly focus on mobile positioning only. \cite{wildfire} presented an interesting application adopting satellite images and linked geographic data to detect wildfire. Ontology and RDF stream processing had also been adopted to develop autonomous vehicles in \cite{trafficAccident}. A novel approach for spatiotemporal query linked data was reported in \cite{spatioQuery}. \par
Scholte and Rozenkrane \cite{385} have proposed a system to localize and track each ship, and send personalized alert to those that are expected to be in danger. However, this system might lose function when the communication device fails, and cannot detect an accident when the communication device on the ship is not functioning. \cite{386} designed and developed an ontology for emergency notification, such as ``a typhoon approaching'', but not for detection tragedies already happened.  

\section{Conclusions}
\label{sec_conclusion}
In this paper we have shown that the system data inside cellular networks can be used to detect accidents with large number of phones lost signal at the same location simultaneously, in our scenario a ship capsizing incident. 
\par
In lost silence, heterogenous user data from all the vendors are concentrated and organized with Geo-Pixel, which is proposed as a geography unit. Continuous RDF processing is adopted to query the data streams in real time. We perform the simulation based on a real life shipwreck incident in China 2015, and discussed the results in different scenarios.
\par
This lost silence service adopts the ontology developed and built for the TOUCAN project. Current ontologies of telecommunication networks generally do not provide information and knowledge inside the system. In our future work, we will intent to extend the lost silence with machine learning approaches and real data from cellular networks. 

\section*{Acknowledgements} 

This research was supported by the EPSRC TOUCAN project (Grant No. EP/L020009/1).


\end{document}